\documentclass[aps]{revtex4}
\begin{document}
\draft
\preprint{}

\title{{ \sl Note: this is the original unchanged 1993 (preprint) text of the hard-to-find article
published in: }\\
Proceedings of the International School of Physics ``Enrico Fermi'', 
Course CXXI:
{\it ``Perspectives in Many-Particle Physics''},\\
{\sl eds. R. Broglia and J. R. Schrieffer (North Holland, Amsterdam 1994) pp 5-30}. 
\\ \bigskip  ~ \\
Luttinger's Theorem and Bosonization of the Fermi Surface}
\author{F. D. M. Haldane}
\address{Department of Physics, Princeton University - Princeton NJ 08544}

\begin{abstract}
A Course of Four Lectures given at the
INTERNATIONAL SCHOOL OF PHYSICS 
$\langle\langle$ ENRICO FERMI $\rangle \rangle $,
Varenna on Lake Como, Villa Monastero, Italy, July 1992.
\end{abstract}
\maketitle
\section{Introduction}

In the usual approach to interacting fermion systems, 
the starting point is the treatment of the ideal Fermi gas, followed
by the development of diagrammatic perturbation theory.   Finally,
contact with the phenomenological Landau Fermi-liquid theory is made,
with a discussion of Landau quasiparticles and collective modes.
An exception to this paradigm is found in one-dimensional
systems, where perturbation theory diverges, and the quasi-particle
structure is in general destroyed by  interactions.   In this case,
a different paradigm has developed from Tomonaga's observation
that the low-energy degrees of freedom of a 1D Fermi gas are completely
collective, and the development of the ``bosonization'' technique.
In these lectures, I will try to present a generalization of the 
``bosonization'' description as a general treatment of Fermi surface
dynamics in {\it any} dimension.   This suggests some new interpretations
of the Fermi surface as an ``order parameter'' for metals, and of its
notional formation as $T \rightarrow 0 $ as a type of critical phenomenon.
One virtue of a treatment that starts with the collective description,
arriving at quasiparticles (for $ d > 1$) at the end, is that the special
features that distinguish $d=1$ and $d > 1 $ occur at the end of the treatment,
rather than right at the beginning.
It also makes clear that the existence of a Fermi surface is not necessarily
synonymous with the validity of the Landau quasiparticle description.
It seems in principle possible that systems with a Fermi surface but which
are non Landau Fermi-liquids may exist, and the bosonization methods
seem promising tools for investigating such possibilities.

I will start by developing a one-dimensional interpretation of
bosonization as Fermi-surface dynamics, then extend it to higher
dimensions, review spin-charge separation and fractionalization of
electrons into spin and charge degrees of freedom, and end with
some intriguing new results on persistence of special features
of the ideal gas in some solvable models with ``spinons''.

\section{Luttinger's Theorem}

I will take the conceptual starting point for the bosonization of
the Fermi surface to be the Luttinger theorem\cite{luttinger} expressing
the total particle number and momentum of the Fermi gas  purely
in terms of the
Fermi surface geometry.  
I will initially describe the treatment of one-dimensional,
spinless fermions, and eventually extend it to three-dimensional
electrons with spin. 

The particle and momentum density of a one-dimensional Fermi gas are given by
$$ {2\pi N /L} = \int dk \, n(k) $$
$$ {2 \pi P / \hbar L } = \int dk \, k n(k) $$
where for  free fermions at zero temperature
$$ n(k) = \theta ( E_F - \epsilon(k) ) . $$
We must now express this in terms of the Fermi surface geometry.
In this case, the Fermi surface is described by a set of Fermi points
$\{k_{Fi}\}$ at which there is a step discontinuity 
$\Delta \nu_i = n(k_{Fi}+\delta)
- n(k_{Fi}-\delta)$ in $n(k)$, with $\Delta \nu_i $ = $\pm 1 $ and $\sum_i
\Delta \nu _i $ = 0.  We may then write 
$$ {2 \pi N / L} = \sum_i \Delta \nu_i k_{Fi} , $$
$$ {2 \pi P / \hbar L } = \frac {1}{2} \sum_i \Delta \nu_i (k_{Fi})^2 . $$
For free electrons this is a ``trivial'' result.  However
the  deep result of Luttinger is that (with some reinterpretation) this
result remains valid (at least in perturbation theory) even when there are
interactions between the fermions. In this case $\Delta \nu $ is no longer
the value of a step discontinuity in $n(k)$: $k_{Fi}$ still marks a
singularity in $n(k)$, but (in one dimension) it is generally weaker
than a step singularity.  Instead, the absolute value of $\Delta \nu $
is an index characterizing the nature of the Fermi surface singularity,
and its sign characterizes the {\it orientation} of the surface, which
in one dimension has an outward normal pointing either to the
right or to the left.   The usual value $|\Delta \nu | = 1 $ indicates that
the singularity in $n(k)$ arises from the Pauli principle, but other
rational values such as $1/m$ can occur, 
principally in connection with the fractional
quantum Hall effect, so I will develop the treatment for general
$\Delta \nu$.  

Luttinger's theorem is proved using methods of diagrammatic perturbation
theory, which in any case fails to converge in one-dimensional systems.
It follows from the fact that particle number and momentum are additively
conserved quantities carried by the particles and conserved in total
at each interaction vertex of a diagram.  Since I am invoking Luttinger's
theorem outside the strict validity of its derivation from diagrammatic
perturbation theory, I am in essence taking it as an axiom that is in
principle justified by experimental fact.  
The aim of this treatment will be  
to show it can be taken as the starting point
for the discussion of Fermi surface dynamics.

One further comment is in order.  
The Luttinger theorem for the total momentum assumes strict momentum
conservation; on a lattice, momentum is only conserved modulo
reciprocal lattice vectors.  However, unless the Fermi surface geometry
is commensurate with reciprocal lattice vectors, Umklapp processes are ``frozen
out'' at low temperature, and the non-conservation of momentum on a
lattice is technically an irrelevant perturbation to the low-energy
fixed point.

To proceed, I now formulate the Luttinger theorem in a differential,
{\it local} form. On lengthscales $ \xi $ where
$$ |k_{Fi} - k_{Fj}| \xi >> 1 \quad (i \ne j ) $$
we can {\it locally } define the Fermi surface $k_{Fi}(x)$.  Low-energy,
long-wavelength excited states will then be described purely in
terms of local Fermi surface fluctuations about the ``reference''
(ground state) Fermi surface $k_{Fi}^0$:
$$ k_{Fi}(x,t) = k_{Fi}^0 + \delta k_{Fi}(x,t) . $$     
This is essentially a ``semiclassical'' treatment of the Fermi surface
where momentum and position are simultaneously specified on
a coarse-grained scale.
The local charge density $ \rho (x) $ (relative to the uniform
density ground state) is 
then given by
$$ 2 \pi  \rho (x) = \sum_i \delta \nu_i \delta k_{Fi}. $$
Similarly, the  local momentum density $ \Pi (x) $ is
$$ 2 \pi \hbar^{-1} \Pi (x) = \sum_i \delta \nu_i \left (
k_{Fi}^0 \delta k_{Fi}(x) + \frac {1}{2} (\delta k_{Fi}(x))^2 \right ) .$$

Thus the generators of continuous symmetries (particle conservation,
or $U(1)$ gauge invariance, and translations) are expressed purely in terms
of the locations of the $T=0$ singularities of $n(k)$, now defined locally
on large lengthscales.  The quantities $\Delta \nu _i $ are ``adiabatic
invariants'' that remain unchanged as the  Hamiltonian is adiabatically
varied (and the $k_{Fi}^0$ in general change), provided the basic structure
of the Fermi surface does not change, and are defined by the differential
relation
$$ \delta \rho (x) / \delta k_{Fi}(x') 
= (2 \pi )^{-1} \Delta \nu_i \delta (x-x') . $$
I again stress that the Pauli principle gives $\Delta \nu = \pm 1 $, and
while the unit step-discontinuity in the free-fermion $n(k)$ is
reduced to a step $Z < 1 $ in Landau Fermi-liquid theory, and a weaker
power-law discontinuity in the 1D Luttinger-liquid, the value of
$\Delta \nu $ which characterizes the Luttinger theorem remains fixed. 

The case $\Delta \nu \ne \pm 1 $ occurs in the application to edge states
in the fractional quantum Hall effect (FQHE), which have been extensively
described by Wen\cite{wen}.   As a thought experiment, put  electrons
in a strong uniform magnetic field in the $z$-direction and confine
them to the $xy$-plane and the lowest Landau level.   
Add a substrate potential 
$V(y) $ that is translationally-invariant in the $x$-direction.  The
single-particle dispersion relation is
$$ \epsilon (k_x) \approx \frac {1}{2} \hbar \omega_c + V(k_x\ell^2) $$
where $\ell = (\hbar c /e B)^{1/2}$ is the ``magnetic length'', and 
$V(y)$ is assumed to vary slowly on this scale.  

In this geometry
the $N$-particle Laughlin state takes the form\cite{thouless}
$$ \Psi_m \propto \prod_{i<j}(z_i-z_j)^m \prod_i (z_i^Je^{-y_i^2/2\ell^2}) $$
where $z_j = \exp (i (x_j + iy_j)\ell) $.  This state has non-zero
$n(k)$ in the range 
$$2\pi J/L = k_{F-} < k < k_{F+} =  2\pi (J + m (N-1)) /L ,$$
and the mean
occupation of states in this range is $1/m $. A caricature of this state
is given by occupying $N$ orbitals in this range so that $m-1$ empty
orbitals separate successive occupied orbitals, giving the mean occupation
$1/m$.   For $m=3$, the resulting occupation pattern is the binary string
$\ldots 1001001001001 \ldots $.

Such a state, interpreted as a Slater determinant, is the
Tao-Thouless state\cite{tt}, advanced as a rival model to the Laughlin state
in the early days following discovery of the FQHE.  Taken literally, this
state is not a good model for the FQHE, but the Tao-Thouless configuration
is in a real sense the ``root configuration'' of the Laughlin state:
if it is acted on by the projection operator on the Hilbert subspace of
wavefunctions that vanish as $(z_i-z_j)^m$ as any pair of particles
approach each other, the Laughlin state results.   The only
occupation number configurations contained in the Laughlin state are
those which can be obtained from the root Tao-Thouless state by a succession
of ``squeezing'' operations where a pair of occupied orbitals $k_1, k_2$
are replaced by $k_1', k_2'$ where $k_1+k_2 = k_1'+k_2'$ and
$k_1<k_1'<k_2'<k_2 $.

The bosonization treatment shows\cite{wen} that the occupation
number distribution $n(k)$  of the Laughlin state
vanishes as $|k-k_{Fi}|^{m-1}$ as the ``Fermi points'' are approached
from the interior of the occupied region.  The width of the
occupied region in $k$-space is $2\pi m \rho $ so the generalized
Luttinger theorem states that
$$  \delta \rho (x) / \delta k_{F\pm}(x')
= \mp (2 \pi m)^{-1} \delta (x-x') . $$
Since the {\it mean} value of $n(k)$ in the interior region is $1/m$, but
(for $m > 1$) $n(k)$ must be less than this near the edges, there
will be regions where $n(k)$ exceeds $1/m$.  
(Numerical studies\cite{halrezayi} confirm that, going into the
interior of a wide strip of Laughlin state, $n(k)$ rises to 
to a maximum, the oscillates as it relaxes to its uniform value.)
If $(k_{F+}-k_{F-})\ell >> 1 $
the edges are well separated, and the deviations from the mean
occupation $1/m$ will be localized in the edge regions.
In this case, if one edge moves as charge is added, it is easy to
see that the Luttinger theorem is satisfied, as the non-interacting
edges will preserve their shape as they move, and the ``extra occupation''
will go into the uniform interior region of the $n(k)$ distribution.  However,
the principle is still valid if the edges are spatially close and there is
{\it no} uniform density interior; in this case, the shape of the
$n(k)$ distribution will deform so as to satisfy the sum rule.
We thus see that the Luttinger theorem can have non-trivial non-Pauli-principle
extensions.

The total conserved quantities are thus written
$$ \Delta N = \sum_i \Delta N_i $$
where
$$ \Delta N_i = \int {dx \over 2 \pi } \Delta \nu_i \delta k_{Fi}(x) , $$
and
$$ \Delta P = \hbar \sum_i k_{Fi}^0\Delta N_i +
\frac{1}{2} \int {dx \over 2 \pi } \Delta \nu_i
(\delta k_{Fi})^2. $$
The crucial feature is that at the metallic $T = 0$ fixed point of the
system
the charges $\Delta N_i $ are {\it separately conserved}
as a consequence of momentum conservation at low energies.
In the microscopic Hamiltonian, of course, only the total charge
$\Delta N$ is conserved.  It is important to note that the Fermi
surface (in the sense of a singularity in the $n(k)$ distribution)
strictly only appears in the $T \rightarrow 0 $ limit, and it is only
in this limit that the Luttinger theorem becomes precise.  The existence
of the Fermi surface at $T = 0 $ {\it implies a separate charge conservation
law ($U(1)$ gauge symmetry) at each point on the Fermi surface}.  In some
sense, this is just a restatement of the Landau Fermi-liquid theory
principle that the lifetime of a Fermi-liquid quasiparticle becomes infinite
at the Fermi surface as $T \rightarrow 0 $.  From 
this viewpoint, the formation of the Fermi surface as $T \rightarrow 0$
is a critical phenomenon, and it is not surprising that new symmetries
not present in the microscopic Hamiltonian appear at a critical point:
in a renormalization-group sense, 
the symmetry-breaking terms present in the microscopic model will
correspond to irrelevant perturbations of the fixed-point effective
Hamiltonian.

We must now construct the most general effective Hamiltonian compatible
with separate charge conservation at each Fermi point.  For free
electrons, with $\Delta \nu = \pm 1 $, the Hamiltonian 
$\Delta {\cal H}$ = $\Delta (H - E_F N)$ is given by
$$ \Delta {\cal H}_0 =  \frac{1}{2} \sum_i \int {dx \over 2\pi} 
v_{Fi}( \delta k_{Fi}(x) )^2  $$
where the Fermi velocity is $v_{Fi}\Delta \nu_i $. (If charge is added
to change $k_{Fi}$ by $\delta k_{Fi}$ the mean momentum of the additional
particles is $k_{Fi}^0 + \frac {1}{2} \delta k_{Fi} $.)  The most general
possibility is essentially a Landau-type form
$$ \Delta {\cal H}^{eff} = 
 \frac {1}{2} \sum_{ij} \int {dx \over 2 \pi } \int {dx' \over 2\pi }
\Gamma_{ij}(x-x') \delta k_{Fi}\delta k_{Fj} . $$
Here stability requires that
$$ \tilde{\Gamma}_{ij}(q) = \int dx \, \Gamma_{ij}(x) e^{iqx} $$
is a real positive-definite symmetric matrix.

It is useful to review the conservation laws and gauge symmetries of 
(a)  of free electrons,
(b)  of the microscopic Hamiltonian,
(c) of this effective Hamiltonian, and
(d) of the Landau Fermi liquid ($D \ge 2$).
Free spinless fermions have a huge set of $U(1)$ gauge symmetries, 
one for each
orbital with a conserved occupation number
in the $D$-dimensional reciprocal space ({\it i.e.}, 
there is a $D$-dimensional
manifold of gauge symmetries). 

In contrast, only the global $U(1)$ symmetry is present in the microscopic
interacting model.  The degrees of freedom in the effective Hamiltonian
derived above correspond only to fermion orbitals close to the Fermi surface;
its explicit symmetries correspond to gauge changes where all orbitals directly
above and below
a given point on the Fermi surface have the same phase change, corresponding
to a $(D-1)$-dimensional manifold of gauge symmetries (one for each point
on the Fermi surface). (In 1D this is a discrete set, one per Fermi point.)

Finally, because the interactions in a Landau Fermi liquid leave a
low-energy spectrum of fermionic quasiparticles in one-to-one correspondence
with bare electron states ``near'' the Fermi surface, it has the
full $D$-dimensional manifold of gauge symmetries asymptotically close
to the Fermi surface, with conserved quasiparticle occupation numbers.

{From} this we conclude that the {\it existence} of the Fermi Surface
implies only the $(D-1)$-dimensional manifold of symmetries, not the
full $D$-dimensional manifold that reappears in the Landau theory.
This indicates that the Fermi surface can exist even when the Landau
quasiparticle picture is not applicable, and non-Fermi-liquid systems
can still have a Fermi surface obeying the Luttinger theorems.  This
is in fact the case in interacting 1D systems (which are not
Landau Fermi liquids), and leaves open the
possibility (discussed later) of such a possibility for $D > 1$.

The term ``Luttinger liquid''\cite{pwa} has been used by Anderson
to refer to a system with a Fermi surface that
obeys the Luttinger theorem, but which is not a Landau
Fermi-liquid.  As a historical note, when I coined the
term ``Luttinger liquid'' in the 1D context\cite{haldane81} I was referring
to  the exactly solvable Luttinger model\cite{lutmodel,liebmattis}
which in 1D  played the role of the ``zeroth order'' model to which residual
interactions are added, in analogy to Landau's use of the
free fermion model as the ``zeroth-order'' model for $D \ge 2$ Fermi liquids. 
However,
it is serendipitous that Luttinger originated  both the theorem and the
model, and Anderson's interpretation of the term 
``Luttinger liquid'' is particularly appropriate.

\section{Quantization of the Fermi-Surface variables.}
So far, I have derived expressions for $\Delta {\cal H}$, $\Delta P $, and
$\Delta N$ relative to the ground state as quadratic expressions in
terms of the local Fermi-surface displacements $\delta k_{Fi} (x) $.
To quantize these degrees of freedom, we need to find their
dynamical algebra.   This is found by the Tomonaga's {\it bosonization}
method\cite{tomonaga}.  

The {\it total} electron density $\rho (x) $ has commuting Fourier
components $[\rho_q,\rho_{q'}] = 0 $, where
$$ \rho_q = \sum_{kk'} \delta_{k-k',q} (c^{\dagger}_kc_{k'} - 
\langle 0 |c^{\dagger}_kc_{k'} |0 \rangle ) $$
The Tomonaga procedure is to decompose the total electron density into
components associated with each Fermi point:
$$ \rho_q \approx \sum_i \rho_{qi} .$$
This is done by defining small non-overlapping domains of width $\Lambda $ in
reciprocal space around each Fermi point:
$$ f_i(k) = \theta (\Lambda^2 - |k-k_{Fi}^0|^2 ) $$ 
Then
$$ \rho_{qi} = \sum_{kk'} f_i(k)f_i(k')\delta_{k-k',q} 
(c^{\dagger}_kc_{k'} - \langle 0 |c^{\dagger}_kc_{k'} |0 \rangle ) .$$
The commutation relations are
$$[\rho_{qi},\rho_{q'j}] = \delta_{ij}\left 
( \delta_{q+q',0} \sum_k f_i(k)f_i(k+q)
\langle 0 | n_{k+q} - n_{k} | 0 \rangle + X_i(q,q') \right ) ,$$ 
where
$$ X_i(q,q')  =  \sum_{kk'} f_i(k)f_i(k') [ f_i(k'+q)-f_i(k'+q') ]
(c^{\dagger}_kc_{k'} - \langle 0 |c^{\dagger}_kc_{k'} |0 \rangle ) .$$
For $|q| << \Lambda $, the factor $f_i(k)f_i(k+q) \approx 1 $ over
almost all the range where $\langle 0 | n_{k+q} - n_k | 0 \rangle $ is
non-negligible.  Similarly, the factor $f_i(k)f_i(k')[ f_i(k'+q)-f_i(k'+q') ]$
vanishes over most of the range of $k$ and $k'$.  Up to corrections of order
$|q|/\Lambda $, the commutator becomes
$$ [\rho_{qi},\rho_{q'j}] = \delta_{ij}\delta_{q+q',0} (qL/2\pi )
\Delta \nu_i ,$$
where $\Delta \nu_i  $  is the shift in $\langle 0 | n_k | 0 \rangle $ in 
going from $k \approx k_{Fi} - \Lambda $ to $k \approx k_{Fi} + \Lambda $.
With the identification $2\pi \rho_i (x) = \Delta \nu_i \delta k_{Fi}(x) $,
we obtain the local form of the
dynamical algebra of the Fermi-surface dispacements: taking
$\Delta \nu _i $ to be the rational number  $\xi_i p_i /q_i $
with $p_i $ and  $q_i$ positive, and $\xi_i = \pm 1 $,
$$ [\delta k_{Fi}(x) , \delta k_{Fj} ] = (2 \pi i q \xi_i /p )\delta_{ij} 
\delta '(x-x') . $$
(Here $\delta ' (x) $ is the derivative of the Dirac delta-function.)
Since the RHS of this commutation relation is a c-number, and the
Hamiltonian is quadratic, the effective Hamiltonian has been  reduced
to a harmonic oscillator problem.
 
The operator that creates an electron in a wave-packet of states
near the $i'th$ Fermi point has the form
$$ \Psi^{\dagger}_i(x) = A_i \exp \left ( i \int^x dx' (k_{Fi}^0 +
\delta k_{Fi}(x) \right ) .  $$
We may write this as
$$ \Psi^{\dagger}_i (x) = A_i e^{ i\varphi_i(x) } .$$
It changes the total charge at the Fermi point by one unit:
$$ [\Delta N_i , \Psi^{\dagger}_j (x) ] \ \delta_{ij} \Psi_i^{\dagger} (x) .$$
{From} the Tomonaga commutation relations,
$$ [\varphi_i(x), \rho_j(x') ] = i\delta (x-x') $$
so $\varphi_i(x)$ is the conjugate field to $\rho_i(x)$, and  is
subject to the {\it chiral constraint}
$$ \partial \varphi_i(x) / \partial x = k_{Fi}^0+  \delta k_{Fi}(x) 
= k_{Fi}^0 + 2 \pi i (\xi_i q_i /p_i) \rho_i(x).$$
Thus $\rho_i (x) $ and $\varphi_i (x)$ are not independent canonical
fields, and $\rho_i(x)$ is proportional to the derivative of its own
conjugate field.
The explicit representation of $\varphi_i(x)$ is
\begin{equation}
\varphi_i(x) = \theta_i + \int_0^x dx' {\partial \over  \partial x' } 
\varphi_i(x) .
\end{equation}
The integration constant $\theta_i$ is the conjugate phase to
the number operator $\Delta N_i$, and is {\it not}` constructed from the
hamonic fluctuation modes.

Integrating the commutation relations gives
$$ [ \varphi_i (x), \varphi_j(x') ] = (i \pi \xi_i q_i /p_i ) \delta_{ij}
{\rm sign }( x-x') . $$ 
{From} this we get (for $x \ne x'$)
$$ \Psi_i^{\dagger} (x) \Psi^{\dagger}_i(x')
= e^{i\pi q_i /p_i } 
\Psi_i^{\dagger} (x') \Psi^{\dagger}_i(x). $$
For $\Psi_i^{\dagger}(x) $ to be a fermion creation operator, we
require that $p_i = 1 $ and that $q_i$ is {\it odd}.
Anticommutation of operators $\Psi_i^{\dagger}(x)$ and 
$\Psi_j^{\dagger}(x')$ when $i$ and $j$ are different requires that
the {\it Klein factor} is
$$ A_i = \exp \left (  i (\pi /2)\sum_j {\rm sign} (i-j) \Delta N_j \right ) $$
where an arbitrary ordering of the Fermi points has been introduced. 

\section{Diagonalizing the One-Dimensional Hamiltonian}

Let us first consider systems with no coupling between different
Fermi points, so $\Gamma_{ij}(x-x') = \gamma_i \delta_{ij} \delta (x-x')$,
with $\gamma_i > 0 $.  When $\Delta \nu = \pm 1 $, this 
just describes free electrons, or Landau quasiparticles.  I will extend
the discussion to the Laughlin-state case $\Delta \nu = \pm 1/m$, following
WEN \cite{wen}, as this is relevant for fractional 
quantum Hall effect edge states.

The Hamiltonian is a sum of decoupled terms associated with each
Fermi point, ${\cal H}$ = $\sum_i {\cal H}_i $,where
\begin{equation}
{\cal H}_i = \epsilon_F\Delta N_i + \frac{1}{2} \int dx \, \gamma_i
: (\rho_i(x))^2 :
\end{equation}
Using the commutation relation $[\rho_i(x),\rho_i(x')]$ =
$ i (2\pi m)^{-1}\xi\delta'(x-x')$, with $\xi = \pm 1$,
we get 
\begin{equation}
[{\cal H},\rho_i(x) ] = i\xi_i v_{Fi} {\partial \over \partial x } \rho_i(x)
\end{equation}
where the Fermi velocity is given by $v_{Fi} = \gamma_i /(2 \pi m) $.
In terms of boson creation and destruction operators obtained by normalizing
the Fourier components of $\rho_i(x)$, the Hamiltonian is given by
\begin{equation}
{\cal H}_i = \epsilon_{Fi} \Delta N_i + v_{Fi}
\left ( \frac{1}{2} {2 \pi m \over L } (\Delta N_i)^2
+ \sum_q \theta (q \xi_i) q b^{\dagger}_qb_q  \right ).
\end{equation}
The $m$-dependence shows up only in the term involving the total charge
at the Fermi point.   It is straightforward to compute the Greens function
using the bosonic representation of the fermion creation operators:
\begin{equation}
\langle \psi_i(x,t)\psi_i{\dagger}(0,0) \rangle 
\approx {Z_i e^{i(k_{Fi}x-\epsilon_{Fi}t)} \over (x-v_{Fi}\xi_it)^m }
\end{equation}
where $Z_i$ is an undetermined normalization.
The Fourier transform gives the singular part of the occupation factor
$n(k)$:
\begin{equation}
n(k) = n(k)_{reg} + Z_i |k-k_{Fi}|^{-1} (k-k_{Fi})^m .
\end{equation}
More generally, the electron creation operator will take the form
\begin{equation}
\psi^{\dagger}(x) = \sum_{\{n_i\}} A(\{n_i\}) e^{i \sum_i n_i \varphi_i(x)} ,
\end{equation}
where $A(\{n_i\})$ = 0 unless $\sum_i n_i $ = 1.   The occupation
factor $n(k)$ will in general have singularities at
$k$ = $\sum_i n_i k_{Fi} $.  

I now turn to the full problem, when $\Gamma_{ij}(x)$ 
$\approx$ $\Gamma_{ij} \delta (x-x') $is not diagonal, and the
different Fermi points are coupled.   The problem is to diagonalize
\begin{equation}
{\cal H} = {2 \pi \over 2 L } \sum_{ij} \Gamma_{ij} \rho_{qi}\rho_{-qj}
\end{equation}
\begin{equation}
[\rho_{qi},\rho_{q'j} ] = (qL/2\pi) \delta_{q+q',0}(\delta_{ij}\xi_i /m_i ) .
\end{equation}
The normal modes $\rho^{\lambda}_q $
are obtained from a real non-symmetric matrix eigenproblem:
\begin{equation}
\rho^{\lambda}_q = \sum_i \chi^{\lambda}_i \rho_{qi} ;
\quad \rho_{qi} = \sum_{\lambda}\psi^{\lambda}_i \rho^{\lambda}_q .
\end{equation}
where
\begin{equation}
\sum_i \chi^{\lambda}_i\psi^{\lambda '}_i
 = \delta^{\lambda\lambda '} ;
\quad \sum_{\lambda} \chi^{\lambda}_i \psi^{\lambda}_j = \delta_{ij} .
\end{equation}
and
\begin{equation}
\psi^{\lambda}_i = (\Gamma^{-1})_{ij} \chi^{\lambda}_j ;\quad
(\xi_j/m_j) \chi^{\lambda}_j = v^{\lambda}\psi^{\lambda}_j .
\end{equation}
Here, the eigenvalues $v_j$ is real, since $\Gamma_{ij}$ is positive definite;
these are the set of renormalized normal-mode velocities.  
To calculate the Greens' function, for example,  the expression for
$\rho_{qi}$ in terms of the  normal modes $\rho_q^{\lambda}$
must be substituted into $\varphi_i(x)$.  The result for the diagonal
Greens function at Fermi point $i$ is
\begin{equation}
G_i(x,t) = Z_i \prod_{\lambda} (x-v^{\lambda}t)^{-\alpha_{i\lambda}},
\end{equation}
where $\alpha_{i\lambda} $ = $(\chi^{\lambda}_i)^2/2\pi |v_{\lambda}| $.
Detailed examination shows that $\sum_{\lambda} \alpha_{i\lambda}  \ge m_i $,
so the Fermi surface singularity in $n(k)$ is always {\it weakened} by
coupling between different Fermi points.   General expressions for
correlation exponents when many different Fermi points interact
have also been developed by PENC and SOLYOM\cite{penc}.

I now come to what is one of  the central ideas of the ``Luttinger
liquid theory''\cite{haldane81}.   This is that (unless crossover
to a non-Luttinger-liquid fixed point occurs), the Landau parameters
$\Gamma_{ij}$ can be determined by identifying the excitations
of a finite interacting system with periodic boundary conditions,
that are associated with changing the net charges at the different
Fermi points.   If we suppress the finite-wavelength harmonic
oscillator modes, the residual charge terms in the excitation
spectrum are:
\begin{equation}
\Delta P = \sum_i k_{Fi}^0\Delta N_i + {\pi \over L} \xi_im_i (\Delta N_i)^2 ,
\end{equation}
\begin{equation}
\Delta {\cal H}  = \sum_i \epsilon_{Fi}\Delta N_i + {\pi \over L}
\sum_{ij} \Gamma_{ij} \Delta N_i \Delta N_j .
\end{equation}
By fitting the low-energy excitations of a system studies by finite-size
numerical diagonalization or the Bethe Ansatz to this form, the low-energy
effective Hamiltonian is determined, and its asymptotic correlations, 
{\it etc.} can be calculated from them.   The program was demonstrated
in detail on the spinless fermion system  equivalent to the XXZ spin chain
in a magnetic field\cite{haldane80}, where there are just two Fermi
points, R and L.  The correspondence between the parametrization
of \cite{haldane80} and that used here is that
$\Gamma_RR$ = $\Gamma_LL$ = $v_N+ v_J$, and $\Gamma_{RL}$ = $\Gamma_{LR}$
= $v_N-v_J$.

Finally, we must discuss what happens when we include electron spin.
If a magnetic field is present, this is just an application of the
``spinless'' treatment with double the number of Fermi points.
However, if the Fermi points have  spin degeneracy, the full
non-Abelian $SU(2)$ symmetry at each Fermi point must be considered.
In this case, coupling between the spin degrees of freedom at different
Fermi points is incompatible with the existence of independent
spin rotation symmetries at each Fermi point.  The renormalization
group treatment\cite{lutheremery} shows that either the couplings scale
to zero at the low-energy fixed point, or they scale to a strong coupling,
non-Luttinger liquid fixed point.

\section{Generalization to Higher Dimensions}

In the one-dimensional systems, the low-energy degrees of freedom
are described by an independent set of bosonic variables at each Fermi
point, representing harmonic fluctuations of the Fermi surface.  
LUTHER\cite{luther} made a pioneering attempt to describe higher-dimensional
Fermi surface degrees of freedom by bosonization, but his ``tomographic''
construction restricts attention to particle-hole pairs carrying
a net momentum strictly normal to the local Fermi vector.
Recently, I found that the Tomonaga bosonization algebra
could be formulated
in a more explicitly higher-dimensional form; this formulation has been
reviewed by HOUGHTON and MARSTON\cite{houghton} 
who use it to discuss corrections
to the $T$-linear specific heat of Fermi liquids.  [Since these lectures
were given, a very similar
treatment has also been independently been developed by 
CASTRO-NIETO and FRADKIN\cite{fradkin}.]

The basic idea is again to remark that the Luttinger theorem expresses
the ground-state particle density and momentum density purely in terms of
the ($d$-1)-dimensional Fermi surface in the $d$-dimensional reciprocal
space defined by the singularity in the ground-state occupation number
distribution $n(k)$.  This of course is making the assumption that
the ground state $n(k)$ has such a feature, and that no BCS or density-wave
instability occurs at low temperatures.  

It is generally believed that
in the absence of any other instability, a BCS instability in some channel
will always occur below some critical temperature,
and destroy the singularity in $n(k)$, so we must in principal
first exclude the BCS processes from the effective Hamiltonian, then
restore the (presumably relevant) perturbations.  The BCS terms can be
recognized as deriving from the special shape of the Fermi surface,
which in the presence of either time-reversal symmetry or spatial
inversion symmetry, has inversion symmetry in reciprocal space.
If, for the moment, we ignore or conceptually abolish this symmetry,
it should in principle be possible to have higher-dimensional interacting
systems with a stable Fermi-surface singularity in their ground state.
{From} this viewpoint the BCS instability, like density-wave
instabilities, is classified as a special feature associated with
a particular class of Fermi surface shapes. 

In the Fermi-liquid theory, the Fermi surface singularity is a
step discontinuity across which $n(k)$ decreases by an amount $Z$,
but I will make no {\it a priori} assumption about the nature of the 
singularity, and merely use the property that it defines a surface
satisfying the Luttinger theorem.   It will become clear that, in
dimensions greater than one, non-Fermi-liquid behavior (such as spin-charge
separation) 
requires a sufficiently-strong singular forward-scattering term in
the phenomenological Landau parameters; the possible existence of such
singular terms, which ANDERSON\cite{pwa} has argued
are generically present in two-dimensional fermion systems, is 
controversial, and 
currently a subject of active
investigation, though to date, no  microscopic treatment has clearly
demonstrated
the existence of such terms. 

In general dimensions, the Fermi surface is described by a function
$\vec{k}_F(s)$, where $s$ is a ($d$-1)-dimensional surface coordinate.
On large lengthscales, I again describe the system in terms of
local fluctuations of the Fermi surface geometry:
\begin{equation}
\vec{k}_F(x,s) = \vec{k}_{F0}(s) + \hat{n}(s)\kappa_{\parallel}(x,s)
+ \hat{t}_{\mu}(s)\kappa_{\perp}^{\mu}(x,s) 
\end{equation}
where $x$ now represents a $d$-dimensional spatial coordinate,  
$\hat{n}(s)$ is the local direction of the Fermi velocity (the outward
normal direction of the Fermi surface), and $\{\hat{t}^{\mu}(s)\}$ are
a basis of the $d-1$ unit vectors tangent to the Fermi surface.   A
treatment that is quadratic in the normal and tangential
fluctuations $\kappa_{\parallel}(x,s),\kappa_{\perp}^{\mu}(x,s)$ 
will be developed, with the recognition that the transverse
fluctuations are essentially gauge variables describing
infinitesimal {\it reparametrizations} of the (curvilinear)
surface coordinates $s \equiv \{s_1,\ldots, s_{d-1}\} $, without
change in the shape of the surface.  The physical
quantities such as the local fluctuation in total particle density
and momentum density can be completely expressed in terms of
the normal fluctuations $\kappa_{\parallel}(x,s)$.   Classically, the
gauge condition $\kappa_{\perp}^{\mu}(x,s) $ = 0 could be imposed;
however, since as a 
quantum operator $\kappa_{\perp}^{\mu}(x,s)$ has non-trivial
commutation relations, the gauge condition is the action
$\kappa_{\perp}^{\mu}(x,s) |\Psi \rangle $ = 0 on physical
states $|\Psi \rangle $.

Thus
\begin{equation}
\Delta N = \int d^{d}x \, \rho (x) 
\end{equation}
where the local change $\rho(x)$ in particle density relative to the
ground state is given by
\begin{equation}
\rho (x) = \int {\omega (s) d^{d-1}s \over (2 \pi ) ^d } \, \kappa_{\parallel}
(x,s) ,
\end{equation}
where $\omega (s)$ is the surface area measure.  Similarly,
\begin{equation}
\Delta \vec{P} = \hbar \int d^dx \, \vec{\Pi} (x) ,
\end{equation}
where
the local change in momentum density is
\begin{equation}
\vec{\Pi} (x) = 
\int {\omega (s) d^{d-1}s \over (2 \pi ) ^d } \, 
\left ( \vec{k}_{F0}(s) \kappa_{\parallel}(x,s)
+ \frac{1}{2} \hat{n}(s) :(\kappa_{\parallel}(x,s))^2 :\right ) .
\end{equation}
(Here the notation $:(\kappa_{\parallel}(x,s))^2 :$ anticipates
the normal-ordering needed in the quantized formulation.)  I note that
the first and second functional derivatives of $\vec{\Pi}(x)$
with respect to $\kappa_{\parallel}(x,s)$ define the two fundamental
geometric properties of the Fermi surface, its {\it shape} 
$\vec{k}_{F0}(s)$ and its {\it orientation} $\hat{n}(s)$.

In the spirit of Landau theory, the effective Hamiltonian is
a quadratic form where $\Delta (H - \epsilon_F N)$ is given by
\begin{equation}
\frac{1}{2} \int d^{d}x \int d^{d}x' 
\int {\omega (s) d^{d-1}s \over (2 \pi ) ^d } 
\int {\omega (s') d^{d-1}s' \over (2 \pi ) ^d } \, 
\Gamma(s,s';x-x') :\kappa_{\parallel}(x,s)\kappa_{\parallel}(x',s'): ,
\end{equation}
where
\begin{equation}
\Gamma(s,s';x) = {(2\pi)^d \over \omega (s) } v_F(s) \delta^{d-1}(s-s')
\delta^d(x-x') + f(\vec{k}_F(s),\vec{k}_F(s');x) .
\end{equation}
Note that the kinetic energy (``effective mass'') term appears as a 
$(2d-1)$-dimensional delta-function term in $\Gamma$. The conventional
Landau $f$-function is given by 
\begin{equation}
f(\vec{k}_F(s),\vec{k}_F(s')) = \int d^d x\, 
f(\vec{k}_F(s),\vec{k}_F(s');x) 
\end{equation}
and $\delta^{d-1}(s-s')$ = 
$\omega (s) \delta^{d-1}(\vec{k}_F(s)-\vec{k}_F(s'))$.
The stability of the Fermi surface against spontaneous shape
deformations requires that
\begin{equation}
\tilde{\Gamma}(s,s';\vec{q} = \int d^dx \, e^{i\vec{q}\cdot \vec{x}}
\Gamma (s,s';x)
\end{equation}
is a positive-definite quadratic form in $s,s'$ for all
$\vec{q}$.

In metals, the Fermi surface will in general consist of a number of
distinct manifolds in the primitive
(or sometimes extended) Brillouin zone.  The formal
integral $\int d^{d-1}s $ can be considered to implicitly include sums over
such discrete band indices distinguishing distinct manifolds.
In the discussion here, I will also assume that the Fermi surface sheets
are smooth differentiable (orientable) manifolds with finite
curvature at all points.  If some microscopic parameter is
varied through a critical point at which the Fermi-surface topology
changes, at the critical point there will be a Van-Hove singularity
on the Fermi surface at which the curvature is infinite, and the
linearized treatment of fluctuations will fail.  Systems at or close to
such critical points may also be a place to look for non-Fermi-liquid
behavior.

Having quadratic expressions for the various conserved quantities
in terms of the fluctuation variables $\kappa_{\parallel}(x,s)$,
we now need the $d$-dimensional version of the quantum 
algebra of the Fermi-surface displacements.   I will first give the
answer, then sketch its derivation using a generalization of
Tomonaga's method.   The commutation relations are
\begin{equation}
[\kappa_{\parallel}(x,s),\kappa_{\parallel}(x',s') ]
= (2\pi )^d i D[\delta^d(x-x')\delta^{d-1}(\vec{k}_F(s)-\vec{k}_F(s'))],
\end{equation}
where
\begin{equation}
D[f(x,s)] \equiv (\hat{n}(s)\cdot \vec{\nabla} ) f(x,s) 
\end{equation}
is a ``covariant derivative''.  The fact that the RHS of the commutation
relation is a $c$-number means 
that $\kappa_{\parallel}(x,s) $ can be expressed as a linear
combination of harmonic oscillator variables.   

As a side comment, I note that if a static magnetic
field $\vec{B}(x)$ is present, the covariant derivative
becomes
\begin{equation}
D[f(x,s)] \equiv (\hat{n}(s)\cdot \vec{\nabla} ) f(x,s) 
 + 2 \pi \Phi_0^{-1} \hat{n}(s) \times \vec{B}(x) \cdot
\hat{t}^{\mu} {\partial \over \partial s^{\mu} } f(x,s) .
\end{equation}
Here $\Phi_0$ is the London flux quantum  $2\pi \hbar /e$.
This derivative encodes the information that if a wave packet of states
centered at real-space position $x$ and Fermi-surface point $s$ is made,
the spatial coordinate evolves in the direction
$\hat {n}(s)$ and the Fermi-surface coordinate evolves in the direction
$\vec{B}(x)\times \hat{n}(s)$.   The {\it rate} at which this evolution
takes place is however encoded in the effective Hamiltonian, rather than
the Fermi-surface displacement algebra.   In what follows, I will
assume that no magnetic field is present.

We may also write $\kappa_{\parallel} (x,s) $ = $D[\varphi (x,s)]$, where
\begin{equation}
[\varphi (x,s) ,\kappa_{\parallel}(x',s') ]
= (2\pi )^d i \delta^d(x-x')\delta^{d-1}(\vec{k}_F(s)-\vec{k}_F(s')) ,
\end{equation}
so $\kappa_{\parallel}$ is the derivative of its own conjugate field.
The scalar phase field $\varphi (x,s) $ obeys the algebra
\begin{equation}
[\varphi (x,s), \varphi (x,s') ] =
i \pi \delta^{d-1}(\vec{k}_F(s) - \vec{k}_F(s'))
\delta^{d-1}(\hat{n}(s)\times (\vec{x} - \vec{x}')) 
{\rm sgn}(\hat{n}(s) \cdot (\vec{x}-\vec{x}')) .
\end{equation}

This allows the commutation relations of both the normal and
tangential Fermi-surface fluctuations to be obtained from the identification
\begin{equation}
\vec{\kappa}(x,s) = \vec{\nabla} \varphi (x,s) .
\end{equation}

The microscopic derivation of the commutation relations follows Tomonaga's
approach to the 1D system.   First, all large momentum-transfer
scattering processes are in principle integrated out, in 
a renormalization-group  sense, leaving an effective Hamiltonian
that keeps only electron states within a reciprocal space distance
$\Lambda $ from the Fermi surface.  The Fermi surface is then broken up into
``patches'' of area about $\lambda^{d-1}$, and reciprocal space
near the Fermi surface is broken up into little domains centered on
each patch.  It is convenient  to consider these domains as little spheres
of radius $\Lambda $ centered on a mesh of points representing a
triangulation of the Fermi surface, but since
we a seeking an effective long-wavelength theory,
the detailed cutoff structure
should not matter.  In this case, the ``patches'' would be circular
($d-1$)-spheres; since a space cannot be fully tiled with non-overlapping
spheres, the radius $\Lambda $ would have to be chosen so that areas
of Fermi surface that are double-counted because  circular patches
overlap are exactly compensated by omitted areas between the patches,
so the sum of all patch areas exactly equals the Fermi surface area.
Other tiling schemes could be used.  An important condition is that $\Lambda$
must be small enough so that within each domain, the Fermi-surface is
quasi-flat.  This means that there must be some finite upper bound
to the Fermi surface curvature, and excludes the possibility of a Van Hove
singularity on the Fermi surface.

Now let $\theta_{\alpha}(\vec{k})$ = 1 if $\vec{k}$ is inside the spherical
domain centered on the patch with label $\alpha$, and let it vanish otherwise.
Then we define
\begin{equation}
\kappa_{\alpha}(\vec{q})
=
\sum_{\vec{k}} \theta_{\alpha}(\vec{k} + \vec{q} ) \theta_{\alpha}(\vec{k})
\left ( c^{\dagger}_{\vec{k}+\vec{q}} c_{\vec{k}} - \delta_{\vec{q},0}
\langle n_{\vec{k}} \rangle _0 \right ) ,
\end{equation}
and
\begin{equation}
\Pi_{\alpha}(\vec{q})
=
\sum_{\vec{k}} \theta_{\alpha}(\vec{k} + \vec{q} ) \theta_{\alpha}(\vec{k})
\hat{n}_{\alpha}\cdot (\vec{k} - \vec{k}_{F\alpha} + \frac{1}{2} \vec{q} )
\left ( c^{\dagger}_{\vec{k}+\vec{q}} c_{\vec{k}} - \delta_{\vec{q},0}
\langle n_{\vec{k}} \rangle _0 \right ) .
\end{equation}
We must now approximately evaluate the commutation relations, and drop
``cutoff-dependent terms'' in the spirit of Tomonaga's treatment.
Then (ignoring any overlap between patches)
\begin{equation}
[\kappa_{\alpha}(\vec{q}),\kappa_{\alpha'}]
= \delta_{\alpha\alpha'} \left (
X_{\alpha}(\vec{q},\vec{q}') +
\delta_{\vec{q}+\vec{q}',0} g_{\alpha}(\vec{q} ) \right )
\end{equation}
where
\begin{equation}
X_{\alpha}(\vec{q},\vec{q}') =
\sum_{k} \theta_{\alpha} (\vec{k} + \vec{q} + \vec{q}' )
\left ( \theta_{\alpha} (\vec {k} +\vec{q} ) - 
\theta_{\alpha}(\vec{k} + \vec{q}') \right ) \left 
(c^{\dagger}_{\vec{k} + \vec{q}
+ \vec{q}' } c_{\vec{k}} - \delta_{\vec{q}+\vec{q}'} 
\langle n_{\vec{k}}\rangle_0 \right  ).
\end{equation}
This operator-valued term can be neglected for 
$|\vec{q}|, |\vec{q}'| \ll \Lambda $, as the factor
$(\theta_{\alpha}(\vec{k} + \vec{q}) - \theta_{\alpha}(\vec{k} +
\vec{q}') )$ vanishes except at the surface of the spherical domain,
and is a ``cutoff-dependent correction''.   Because (in contrast to
the original Tomonaga calculation in 1D) some of this ``correction''
involves states at the Fermi surface, this is perhaps not as
innocuous an approximation in higher dimensions, but appears
to be valid in the long-wavelength limit.   The residual term in the RHS of
commutation relation is the $c$-number term
\begin{equation}
g_{\alpha}(\vec{q}) = \sum_{k} \theta (\vec{k} + \vec{q} )
\theta (\vec{k} \langle (n_{\vec{k} + \vec{q} } - n_{\vec{k} } )\rangle _0 .
\end{equation}
In this case, for $|\vec{q} | \ll \Lambda $ the value of this is just the
number of allowed $k$-space points inside the volume of
reciprocal space swept out by displacing the patch of Fermi surface
by $\vec{q}$.    Note that it is independent of the detailed structure
of $\langle n_{\vec{k}} \rangle _0 $ near the Fermi surface and  only involves
the change in asymptotic values of 
the occupation factor from deep inside to far outside the Fermi surface.  Thus 
the commutation relation becomes
\begin{equation}
[\kappa_{\alpha}(\vec{q} ) ,\kappa_{\beta} (\vec{q}')]
= aV\delta_{\alpha\beta}\delta_{\vec{q}+\vec{q}'.0} 
\hat{n}_{\alpha}\cdot \vec{q} ,
\end{equation}
where $a$ is the surface area of the patch on the Fermi surface.
This a generalization of an Abelian ($U(1)$) Kac-Moody algebra.
The other commutation relations 
are similarly evaluated at long wavelengths
as
\begin{equation}
[\Pi_{\alpha}(\vec{q}),
\Pi_{\beta}(\vec{q}')] =
\hat{n}_{\alpha}\cdot \vec{q} \kappa_{\alpha}(\vec{q}+\vec{q}')\delta_{\alpha\beta}
\end{equation}
and
\begin{equation}
[\Pi_{\alpha}(\vec{q},\Pi_{,\beta}(\vec{q}')]
= \hat{n}_{\alpha}\cdot (\vec{q} - \vec{q}') \delta_{\alpha\beta} 
\Pi_{\alpha} (\vec{q} + \vec{q}') +
\frac{1}{12} \delta_{\alpha\beta} aV \delta_{\vec{q} + \vec{q}' } 
(\hat{n}_{\alpha} \cdot \vec{q} ) ^3 
\end{equation}
The full structure is the generalization of the Kac-Moody and associated
Virasoro algebras to $d > 1$, where at each point on the Fermi surface,
the spatial coordinates separate into one special normal direction
$\hat{n}_{\alpha}$ along which derivatives are taken, and
$d - 1$ transverse directions.  This differs from the earlier
``tomographic'' bosonization
proposed by Luther\cite{luther} which only considers the case when
$\vec{q}$ is parallel to $\hat{n}_{\alpha}$, and does not allow any
natural coupling between Fermi surface points where the normals are
not parallel or antiparallel.  The thermodynamic limit may now be taken,
and Kronecker delta-functions on the discrete mesh of reciprocal
space points allowed by periodic boundary conditions become Dirac
delta-functions.

The fermion phase field $\varphi_{\alpha} (x) $ is formally
given by
\begin{equation}
  \varphi_{\alpha} (x) = \theta_{\alpha} (x_{\perp}) + 
\int_0^{x} dx_{\parallel}'
(\hat{n}_{\alpha}\cdot \vec{\nabla}) \kappa_{\parallel\alpha}(x) ,
\end{equation}
where $\theta_{\alpha} (x_{\perp}) $ is an integration constant.
The operator $\exp ( i \theta_{\alpha} (x_{\perp}) )$ must be interpreted
as the operator that adds charge on the Fermi surface patch $\alpha$.
This will in a wave packet that is completely delocalized along
the direction in real space parallel to $\hat{n}(s)$, but is localized
in the transverse direction to within a distance $\Lambda^{-1}$ of
the transverse  spatial coodinate $x_{\perp}$.   The electron 
creation operator will again be proportional
to $\exp (i \varphi_{\alpha}(x)) $;  as in the 1D case, a Klein factor can be
added to make electron operators defined in different Fermi-surface
patches anticommute.
Electron creation operators defined in the same patch, and sharing
(to within $\Lambda^{-1}$)
a common transverse spatial coordinate $x_{\perp}$ will also automatically
anticommute.       Anticommutation of creation operators in the same
patch, but at different transverse spatial coordinates must be imposed
through the integration constant $\theta_{\alpha} (x_{\perp})$:
\begin{equation}
e^{i\theta_{\alpha}(x_{\perp})}
e^{i\theta_{\alpha}(x_{\perp}')}
+
e^{i\theta_{\alpha}(x_{\perp})'}
e^{i\theta_{\alpha}(x_{\perp})}
 = 0
\quad (| x_{\perp} - x_{\perp}' | \gg \Lambda^{-1} ).
\end{equation}
If there are only one or two transverse dimensions, this can be represented
with Jordan-Wigner or ``anyon'' gauge fields.

It is straightforward to include spin degrees of freedom in the
preceding treatment, and define
\begin{equation}
\kappa_{\alpha\uparrow}(\vec{q}) +
\kappa_{\alpha\downarrow}(\vec{q}) =
2\kappa_{\alpha}(\vec{q}) 
; \quad
\kappa_{\alpha\uparrow}(\vec{q}) -
\kappa_{\alpha\downarrow}(\vec{q}) =
2\sigma^{z}_{\alpha}(\vec{q}) .
\end{equation}
Then
\begin{equation}
[\kappa_{\alpha}(\vec{q} ) ,\kappa_{\beta} (\vec{q}')]
= \frac{1}{2}aV\delta_{\alpha\beta}\delta_{\vec{a}+\vec{q}'.0} 
\hat{n}_{\alpha}\cdot \vec{q} ,
\end{equation}
\begin{equation}
[\sigma^a_{\alpha}(\vec{q} ) ,\sigma^b_{\beta} (\vec{q}')]
= \delta^{ab}\left (
\frac{1}{2}aV\delta_{\alpha\beta}\delta_{\vec{a}+\vec{q}'.0} 
\hat{n}_{\alpha}\cdot \vec{q}  + i \epsilon^{abc}\sigma^c{\alpha}
(\vec{q} + \vec{q}' ) \right ) .
\end{equation}
The spin degrees of freedom now obey a $d > 1$ version of the non-Abelian
$SU(2)$ Kac-Moody algebra.

\section{Diagonalization of the Harmonic Oscillators}

Within the approximation that keeps only the terms which are quadratic in 
the Fermi surface fluctuations, the problem of the interacting Fermi system reduces
to a harmonic oscillator problem.  In fact this is (of course)
essentially just the zero-sound problem of Fermi liquid theory.   It is 
convenient to first rescale the Fermi surface normal fluctuation 
operators, and write
\begin{equation}
\tilde{\rho}_{q\alpha} = \left ( {v_{F\alpha} \over aV } \right ) ^{1/2}
\kappa_{\alpha}(\vec{q}) .
\end{equation}
Then the commutation relations become
\begin{equation}
[\tilde{\rho}_{q\alpha},\tilde{\rho}_{q'\beta}] =
\delta_{\alpha\beta} \delta_{\vec{q} + \vec{q}',0} \omega^0_{\alpha} (\vec{q}),
\end{equation}
where
$\omega^0_{\alpha}(\vec{q}) $ = 
$ v_{F\alpha} (\hat{n}_{\alpha} \cdot \vec{q} )$.
The Hamiltonian is now 
\begin{equation}
\frac{1}{2} \sum_{q} \sum_{\alpha\beta}\tilde{\Gamma}_{\alpha\beta} (\vec{q})
\tilde{\rho}_{q\alpha} \tilde{\rho}_{-q\beta} ,
\end{equation}
where $\tilde{\Gamma}_{\alpha\beta}$ is the positive
definite matrix
\begin{equation}
\tilde{\Gamma}_{\alpha\beta}(\vec{q}) = \delta_{\alpha\beta}
+ \Lambda^{d-1} { f(\vec{k}_{F\alpha} , \vec{k}_{F\beta} ; \vec{q})
\over (v_{F\alpha}v_{F\beta})^{1/2}  } ,
\end{equation}
which is an even function of $\vec{q}$.
We now see the reason why the 1D case (with no transverse degrees of
freedom) is special.   {\it In the scaling limit
$\lambda \rightarrow 0$, with  $(d -1) > 0 $, $\tilde {\Gamma}_{\alpha\beta}
\rightarrow \delta_{\alpha\beta}$ (free fermions) }, unless either
(a) $f(\vec{k}_{F\alpha},\vec{k}_{F\beta}) \rightarrow \infty$ or
(b) $v_{F\alpha}v_{F\beta} \rightarrow 0 $.  Put another way, for 
$d > 1$
unless the effective  Landau parameters are singular, the coupling 
between different
patches on the Fermi surface contains a factor (patch area)/(Fermi surface
area) = $1/N_{patch}$, 
and the only modification of the collective excitation spectrum
is that a finite number of zero-sound collective modes are pushed up above
the continuum of modes with frequencies up to $v_F |q| $.

Formally, to diagonalize the harmonic oscillator problem, we must 
express its normal modes $\rho^{\lambda}_q$ in terms of the local modes
$\tilde{\rho}_{q\alpha}$ defined on each patch:
\begin{equation}
\rho^{\lambda}_q = \sum_{\alpha} \chi^{\lambda}_{\alpha} (\vec{q})
\tilde{\rho}_{q\alpha}
\end{equation}
with the inverse relation
\begin{equation}
\tilde{\rho}_{q\alpha} = 
\sum_{\lambda}
\psi^{\lambda}_{\alpha}(\vec{q})\rho^{\lambda}_{\alpha} .
\end{equation}
If this involved the 1D problem of the coupling of a single pair
of Fermi points with opposite-direction normals, this problem would
be simple to treat by expressing it in terms of canonically-normalized
boson creation and annihilation operators, and carrying out a
Bogoliubov transformation.
In the general case, this is not so convenient; instead it can be
recognized (of course) as the zero-sound problem, and regarded as a
real non-symmetric eigenproblem where all the eigenvalues $\omega^{\lambda}$
are real
because the matrix $\tilde{\Gamma}_{\alpha\beta}$ is positive definite.
Then
\begin{equation}
\sum_{\beta} \tilde{\Gamma}_{\alpha\beta}
 \omega^0_{\beta}\chi^{\lambda}_{\beta} 
= \omega^{\lambda}\chi^{\lambda}_{\alpha}
\end{equation}
and
\begin{equation}
\sum_{\beta} \omega^0_{\alpha}
 \tilde{\Gamma}_{\alpha\beta}\psi^{\lambda}_{\beta}
= \omega^{\lambda}\psi^{\lambda}_{\alpha}
\end{equation}
with the orthogonality relation
\begin{equation}
\sum_{\alpha} \psi^{\lambda}_{\alpha}\chi^{\lambda '}_{\alpha} = 
\delta^{\lambda\lambda'} .
\end{equation}
The eigenvalues $\omega^{\lambda}(\vec{q}) $ are real, with the
symmetry $\omega^{\lambda}(-\vec{q})$ = $-\omega^{\lambda}(\vec{q})$.
This formal solution is useful for carrying out calculations of
correlation functions, {\it etc.}

If the Fermi surface is regarded 
as the analog of the ``order parameter''
of a metal, its shape fluctuations are its ``Goldstone modes''.
The cutoff $\Lambda$ means that only modes with $|q| < \Lambda $
should be counted as independent, and there is then one
linear-dispersion ``Goldstone mode'' per patch. These modes
have a spectrum of velocities that becomes continuous in the
in the limit $\Lambda \rightarrow 0 $.  For $\vec{q}$ along some direction
$\hat {\Omega}$, the density of mode velocities 
remains finite at zero frequency
provided that some part of the Fermi surface is tangential to
$\hat{\Omega}$.  It is this feature that gives the universal $T$-linear
specific heat of fermion liquids in this formalism, in contrast to the
$T^d$ specific heat of systems where the Goldstone mode velocities
remain finite.  The $T$-linear heat capacity (or entropy) is extensive
in not only the real-space volume $V$, but also in the Fermi-surface
area, and derives from the (2$d$-1)-dimensional delta-function
term proportional to $\delta^d(x)\delta^{d-1}(\vec{k}_F(s) - \vec{k}_F(s'))$
in $\Gamma (s,s';x)$.  Furthermore, this term controls the upper limit
(the Fermi velocity) to the continuous spectrum of velocities of modes
traveling in a given direction.   {\it In the absence of a contribution
to this delta-function part of $\Gamma$ coming from singular terms in the
Landau parameters, there is no renormalization of the Fermi velocity or
the $T$-linear specific heat}. 

The collective zero-sound modes give  $T^d$ corrections to the
low-temperature specific heat that depend on the Landau parameters.
In fact, in three dimensions, the leading corrections to the 
$T$-linear specific heat is the $T^3\log T$ term elucidated in detail
by PETHICK and CARNEIRO\cite{pethick} within a standard Fermi-liquid approach:
HOUGHTON and MARSTON\cite{houghton} have recently reported that
such terms can also be recovered in the bosonized approach discussed here.

Because of this ``all-or-nothing'' character of the contribution of
the Landau $f(\vec{k}_F(s),\vec{k}_F(s'))$ couplings to a shift in the
Fermi velocity and the $T$-linear specific heat, it is instructive the
consider the case when $f(\vec{k}_F(s),\vec{k}_F(s'))$ is a smooth
function with a strong anomaly in a narrow cone around the
forward scattering direction, and allow some control parameter
to continuously evolve this anomaly into a true delta-function.
If the Fermi surface is spherical, as the cone of the
anomaly becomes narrower,  more terms in the spherical-harmonic
expansion of the Landau parameters are needed to adequately represent
it.   Roughly speaking, there will be one extra zero-sound collective mode
pushed out above the continuum of modes with $0 < \omega < v_F|q|$
for each additional spherical harmonic term that becomes significant.
These modes will proliferate and become dense in the range
$v_F |q| < \omega < (v_F + \delta v_F)|q|$ as the Landau parameters
develop a delta-function singularity in the forward scattering
direction.  Similarly, when the anomaly becomes pronounced, the 
specific heat will develop a ``pseudo-$T$-linear'' regime characterized
by what will become the renormalized Fermi velocity, but which crosses over
to the true unrenormalized $T$-linear regime at lower temperatures; this
crossover temperature vanishes as the singularity in the Landau parameters
develops.

\section{Spin-Charge Separation}

{\it Spin-Charge separation} is seen quite generally in one-dimensional
systems.   It is associated with forward scattering of particles at
the {\it same} Fermi point, and is not directly related to the 
other characteristic one-dimensional phenomenon where the long-wavelength
coupling of the low-energy degrees of freedom of {\it different}
Fermi points renormalizes the correlation function exponents
away from their free fermion values.   In fact, an exactly solvable model,
the ``supersymmetric $t-J$ model'' with inverse-square 
interactions\cite{kuramoto} exists in which spin-charge separation
exists without correlation function exponent renormalizations.
In this case, the simple pole of the electronic Greens function splits
into a branch cut terminated by inverse square-root singularities:
\begin{equation}
{\rm Im.} G(k_F + \delta k,E_F + \delta E) \propto 
\theta ((\delta E- v_s\delta k)(v_c\delta k - \delta E))
\left ( (\delta E- v_s\delta k)(v_c\delta k - \delta E)\right )^{-1/2} .
\end{equation}
An electron injected into the system in a wavepacket of states near 
such a Fermi point, and localized in space will physically separate
into spatially separated charge and spin components, moving with velocities
$v_s$ and $v_c$ as the state evolves.

Could such a phenomenon occur in two dimensions, as proposed
by Anderson\cite{pwa}?
We have seen that in dimensions greater than one, the Fermi velocity is
defined by the upper limit of the continuum of velocities of the 
``Goldstone modes'' (Fermi-surface shape fluctuation modes), and that
this cannot be renormalized by  non-singular Landau couplings.  For
free fermions, and Landau Fermi liquids, the spin and charge velocities
are strictly equal, which as we shall see can be interpreted in terms
of a ``gauge symmetry''.  (The equal spin and charge velocities, defined
by the dispersion relation associated with the low-energy pole of the
Landau Fermi-liquid single-particle Green's function should not
be confused with the propagation velocities of the various spin and
charge fluctuation collective zero-sound excitations that are present
in a Fermi liquid).    To get spin-charge separation in higher dimensions,
singular forward scattering terms that differ in the singlet and triplet
scattering channels would be required, as proposed by ANDERSON\cite{pwa}
in two dimensions.   However, it should again be emphasized that 
his proposal remains controversial.   

The phenomenological
description outlined here treats the Landau parameters as an input,
and cannot provide guidance about their microscopic origin or validity.
It may again be useful to consider what would occur if there was, for
example, a strong forward scattering in the triplet but not the singlet
channel, but not a true singularity.  In this case, at higher energy scales
the spin and charge  degrees of freedom would presumably separate over
shorter lengthscales,  but finally, at the longest lengthscales and
lowest energies, the spin and charge quantum numbers of the electron
would be confined together to form a Landau quasiparticle.  A deconfinement
transition would take place if the Landau parameters were ``tuned''
to become (sufficiently) singular.

While spin-charge separation in two or higher dimensions  remains obscure,
I will now examine it more closely in the one-dimensional context
from a symmetry viewpoint.

\section{Hidden Symmetries in Spin-Charge Separated Systems}

As noted earlier, the ideal Fermi 
gas exhibits an infinite set of gauge symmetries, 
as the occupation numbers
of each orbital are separately conserved.   When spin degrees of freedom
are included, there is a infinite set of non-Abelian $SU(2)$
symmmetries, one for each orbital.   This means that the spin of each
singly-occupied orbital can be independently rotated, the spin 
degeneracy 
of a state with $N$ singly-occupied orbitals is $2^N$, and it is a highly
reducible representation of the global $SU(2)$ group.   When
interactions are ``switched on'' this non-generic structure of the ideal
gas is lost, and the eigenstates will become irreducible
representations of the spin rotation group (assuming no spin-orbit coupling).
The essence of the Landau Fermi-liquid state is that, asymptotically at
the low-energy fixed point, the extra symmetries of the ideal gas are
restored, but in the form of the {\it quasiparticle} occupations.

I now pose the question, if forward scattering processes at 
a Fermi point are included, so as to induce spin-charge separation, but
the other interactions that couple different Fermi points are omitted,
is any remnant of the ``hidden'' quasiparticle gauge symmetries
retained?  There is some remarkable evidence from certain exactly
solvable one-dimensional models that this is indeed the case.
These models are perhaps the closest interacting models to
the ideal gas, and seem to be the simplest non-trivial interacting models.
They have scale-invariant inverse-square interactions, and ground state
wavefunctions which can be considered as (full) 
Gutzwiller projections of free fermion states.

The simplest of these models is the $S=1/2$ spin chain which I and
Shastry introduced independently a few years ago\cite{haldane,shastry}:
\begin{equation}
H = J \sum_{i<j} d(i-j)^{-2} \vec{S}_i \cdot \vec{S}_j
\end{equation} 
Here $d(j)$ = $j$, or $(N\pi ) \sin (\pi k /N ) $ if periodic
boundary conditions on a chain of $N$ sites is used. 
This model only has spin degrees of freedom. but an extension of this
to the ``supersymmetric $t-J$ model'' was introduced by 
KURAMOTO and YOKOYAMA\cite{kuramoto}:
\begin{equation}
H = J\sum_{i<j} d(i-j)^{-2} 
\left (-P_G(\sum_{\sigma}c^{\dagger}_{i\sigma}c_{j\sigma} + h.c. )P_G 
+ (\vec{S}_i\cdot \vec{S}_j - \frac{1}{4} n_in_j \right )
\end{equation}
where $P_G$ is the full Gutzwiller projection operator that prevents
multiple occupancy of any site.  This model has both spin and
charge degrees of freedom, and exhibits spin-charge
separation without coupling of low-energy degrees of freedom at different
Fermi points.  The periodic versions of these models exhibit
remarkable ``supermultiplet'' degeneracies meaning that their energy levels
form highly reducible representations of $SU(2)$.   This is analogous
to the free fermion gas degeneracies, but with a much less straightforward
structure, and is what I will interpret as the remnant of the orbital
occupation number symmetries that survives spin-charge separation.
Since these symmetries just involve the spin sector, it is convenient
to consider just the spin chain.

To put the results into context, it is first useful to consider
the conformal limit, where the low-energy spin degrees of freedom
are described by the $k=1$ Kac-Moody algebra (Wess-Zumino-Witten
conformal field theory\cite{affleck}).  In this language, one
writes a (say, right-moving) spin density field $\sigma^a_{q\alpha}$
associated with a single Fermi point as $J^a_m$, with $q = 2\pi m / L$,
and $m = 0, \pm 1, \pm 2 \ldots $.  Then the Kac-Moody algebra
takes its standard form
\begin{equation}
[J^a_m,J^b_{m'}] = km/2 \delta_{m+m',0} + i\epsilon^{abc}J^c_{m+m'}.
\end{equation}
and the Hamiltonian becomes ${\cal H}^{eff}$ = $v_sP$, where
the momentum $P$ is given by
\begin{equation}
P = {2\pi \over L } L_0 ,
\end{equation}
where
$L_0$ is the ``zero mode'' of the associated Virasoro algebra:
\begin{equation}
L_0 = {1 \over k + 2 } 
\left ( J^a_0J^a_0 + 2 \sum_{m=1}^{\infty} J^a_{-m}J^a_m \right ) 
\end{equation}
where $[L_0,J^a_m] = -m J^a_m $, and $J^a_m |0 \rangle $ = 0 for
$ m > 0$.
The Hamiltonian is very degenerate, since $L_0$ takes only values
$n+h$, where $n = 0,1,2,\ldots $ and (for the $k=1$ SU(2) algebra)
$h = 0 $ for integer total spin, and $h = \frac{1}{4}$ for
half-integral total spin.  The standard descriptions of this spectrum
are through ``Abelian bosonization'' (which is essentially
what has been described
in this these lectures) or the ``Verma module'' basis
(see, {\it e.g.} the book by KAKU\cite{kaku} for an introduction),
but these do not describe the fractional-statistics
particle-like $S=\frac{1}{2}$ excitations
(``spinons'') that turn out to be the appropriate basis for describing the
inverse-square perturbation of the conformal limit.

It is useful to introduce a short-distance ``point-splitting'' cutoff
that regularizes the conformal field theory as follows:
\begin{equation}
{\cal H}^{eff} = \frac{1}{2} \int dx \int dx \,' j(x-x') \vec{\sigma}(x) \cdot
\vec{\sigma}(x') .
\end{equation}
The conformal field theory is recovered in the limit
\begin{equation}
j(x) \rightarrow {v_s \over 4 \pi } \delta (x-x') ,
\end{equation}
or $\tilde {j} (0) $ = $ v_s / 4 \pi $, where
\begin{equation}
\tilde{j} (q) =  \int dx \, j(x) e^{iqx} .
\end{equation}
Usually one takes some point-splitting function $j(x)$ that falls
off exponentially for large separations, but consider the case when
it has algebraic tails falling off as $x^{-(1+\alpha})$. Then as
$|q| \rightarrow 0 $,
\begin{equation}
\tilde{j} (q) = (v_s/4\pi )  + A |q|^{\alpha} + Bq^2 .
\end{equation}
For $0 < \alpha < 2 $, the non-analytic term is the leading correction to the
conformal limit; the new term in the Hamiltonian can be
written
\begin{equation}
H^{(2)} = \sum_{m=1}^{\infty} m^{\alpha} J^a_{-m}J^a_{m}.
\end{equation}
Since this commutes with $L_0$, this term can be studied numerically
by diagonalizing it with the finite-dimensional subspace of states
with a given value of $L_0$.  In the limit $\alpha \rightarrow 0 $,
$H^{(2)} \rightarrow \frac{3}{2} L_0 - \frac{1}{2} J_0(J_0+1) $, where
$J_0$ is the total spin quantum number.    In this limit, $H^{(2)}$ merely
splits the states at a given $L_0$ into groups with the same total spin.
However, for general $\alpha > 0 $, the spectrum of $H^{(2)}$ is completely
broken up into distinct energy levels, each of which
corresponds to an irreducible representation of $SU(2)$ 
with no unexpected additional degeneracies.   This represents the complete
destruction of all the higher symmetries of the conformal
field theory by the point-splitting cutoff.   A striking exception
to this is seen in the special case $\alpha = 1$, corresponding to the
inverse-square fall-off of $j(x)$; in this case the levels partially
regroup into ``supermultiplets'' which a highly reducible representations
of $SU(2)$.   No other ``special'' values of $\alpha$ are detected
by this calculation.

Clearly  a large residual part of the symmetry of the conformal
field theory survives in the presence of the inverse-square
corrections to the conformal limit.  This symmetry has recently
been identified as a ``quantum group'' symmetry\cite{yangian} called the
{\it Yangian}\cite{drinfeld,chari} $Y(sl_2)$, generated by $J^a_0$ and
\begin{equation}
{\cal J}^a = ih\epsilon^{abc}\sum_{m=1}^{\infty} J^b_{-m}J^c_{m} 
\end{equation}
where $h$ is here the ``quantum deformation parameter'' defined
by the ``non-co-commutative co-product'' 
\begin{equation}
\Delta ({\cal J}^a) 
= \openone \otimes {\cal J}^a + {\cal J}^a \otimes \openone
 + \frac{1}{2}ih \epsilon^{abc} J^b_0 \otimes J^c_0 .
\end{equation} 
It is perhaps out of place to describe the technical
aspects of ``quantum groups'' (which are in fact algebras,
not groups) in any detail here; suffice it to say that
quantum groups are infinite-dimensional algebras that are
``quantum deformations'' of Lie algebras, with the feature that they have
a tensor-product operation (the ``co-product'') where (unlike Lie algebras)
the result of a sequence of tensor products depends on the {\it order}
in which they are made (analogous to the action of a sequence of operators
in quantum mechanics).  ``Quantum groups'' are intimately related to
braiding and fractional statistics.  A physical explanation of the 
appearance of quantum groups in connection with spin-charge separation
is that if a spin-1/2 fermion is factorized into independent
spin and charge factors, the two components are each {\it semions},
fractional-statistics entities half-way between fermions and 
bosons\cite{laughlin}.     The Yangian $Y(sl_2)$ is the quantum
group which has $sl_2$ (the Lie Algebra of $SU(2)$ generators) as
a subalgebra.

The discrete spin chain also has this ``quantum group'' symmetry,
$[H,{\cal J}^a]$, with
\begin{equation}
{\cal J}^a = { h \over 2} \sum_{i <j }  \cot (\pi (i-j)/N) 
\epsilon^{abc} S^b_i S^c_j .
\end{equation}
The energy levels are given by\cite{yangian} the construction
\begin{equation}
E = 2J(\pi / N)^2 \sum_{i=1}^M  m_i(m_i - N) ;\quad  e^{iK} = \prod_i \exp
 (2 \pi i m_i /N ) ,
\end{equation}
where $\{m_i\}$ are a set of distinct integers in the range
$0 < m_i < m_{i+1} < N $, subject to the ``generalized Pauli
principle'' that not only are they {\it distinct}, but also that
{\it they cannot be consecutive}.  These quantum numbers
can be represented by a binary sequence of
length $N-1$, where a ``1'' represents a value in the
set $\{m_i\}$. This means that the ground state
sequence is $1010101\ldots 1010101$, and has no consecutive pairs of
zeroes, which represent spinon excitations.  
Removing a ``1'' from the sequence thus creates  a {\it two}-spinon
state $\ldots 101000010101 \ldots$, which can
be rearranged to give states such as $\ldots 1010100101010010101\dots $
A sequence such as this represents a fourfold-degenerate
state with $SU(2)$ representation content $ ( \frac {1}{2} ) \otimes 
(\frac{1}{2})$ = $0 \oplus 1$.  

A sequence 
such as ``\dots 1000001 \ldots'' where there are four {\it successive}
``00'' combinations represents a ``four-string'' in CHARI and PRESSLEY's
\cite{chari} representation theory of $Y(sl_2)$, and hence 
contributes a $S=2$ factor
in the tensor product of $SU(2)$ representations that makes up the
representation of $Y(sl_2)$.  Physically, this was previously
interpreted\cite{fdmh2}
as ``four spinons in the same orbital'' with a selection rule
that spinons ``in the same orbital'' could only be in a symmetric
spin state\cite{stat}.   This empirically-observed rule, discovered
by detailed examination of the results from numerical 
diagonalization\cite{fdmh2}, now is seen to precisely correspond 
to the $Y(sl_2)$ representation theory\cite{chari}.

This  example suggests that ``quantum-group'' techniques may turn out to
have important applications in connection with fractional statistics,
as a more algebraic formulation that makes contact with ``occupation
number'' descriptions and the Pauli principle.

\section{Conclusion}

In these lectures, I sketched out the logic of an
approach to Fermi fluids based on the idea that the
Fermi surface is an analog of an order parameter, and that
the low-energy degrees of freedom can be fully treated in terms
of  ``bosonized'' variables describing local fluctuations
of the shape of the Fermi surface.  The Luttinger theorem
relating the volume of the Fermi surface to the particle
density is seen to be the key principle. 
While bosonization has been a key tool in treating the one
dimensional systems, it clearly shows promise in higher
dimensions too.  Much remains to be done to make this method
a real working tool for  higher dimensions. 
On the one hand, it will be interesting to see how much of
the standard Fermi liquid results can be reproduced 
using such methods.  On the other hand, they seem to have potential 
for the study of possible non-Fermi-liquid states, since they
are not based on a perturbative expansion about the
non-interacting Fermi gas.
I also considered spin-charge separation, primarily in one dimension,
and described some recent hints that ``quantum group'' methods
may be important in cases where fermion variables fractionalize
into fractional-statistics objects.

This work was supported in part by NSF grant DMR 91-96212.

%
%

%
%

\end{document}